\documentclass[12pt]{iopart}
\usepackage{iopams}  
\usepackage{amsmath}
\usepackage{supertabular}

\begin{document}

\title[Reconstruction of Black Hole Metric Perturbations from Weyl Curvature II]
{Reconstruction of Black Hole Metric Perturbations from Weyl Curvature II: The Regge-Wheeler gauge}

\author{Carlos O. Lousto}

\address{Department of Physics and Astronomy,
and Center for Gravitational Wave Astronomy, The University of Texas
at Brownsville, Brownsville, Texas 78520, USA}

\date{\today}

\begin{abstract} 
Perturbation theory of rotating black holes is described in terms of
the Weyl scalars $\psi_4$ and $\psi_0$; each satisfying the
Teukolsky's complex master wave equation with spin $s=\mp2$, and
respectively representing outgoing and ingoing radiation.  We
explicitly construct the metric perturbations out of these Weyl
scalars in the Regge-Wheeler gauge in the nonrotating limit. We
propose a generalization of the Regge-Wheeler gauge for Kerr
background in the Newman-Penrose language, and discuss the approach
for building up the perturbed spacetime of a rotating black hole. We
also provide both-way relationships between waveforms defined in the
metric and curvature approaches in the time domain, also known as the
(inverse-) Chandrasekhar transformations, generalized to include
matter.
\end{abstract}




\section{Introduction}

There is a formulation of the perturbation problem
derived from the Newman-Penrose formalism~\cite{Newman62a} that is
valid for perturbations of rotating black holes.\cite{Teukolsky73}
This formulation fully exploits the null structure of black holes to
decouple the curvature perturbation equations into a single master wave
equation that, in Boyer-Lindquist coordinates $(t,r,\theta,\varphi)$,
can be written as:
\begin{eqnarray}
&&\ \ \Biggr\{\left[ a^2\sin^2\theta -\frac{(r^2+a^2)^2}\Delta \right]
\partial_{tt}-\frac{4Mar}\Delta \partial_{t\varphi }
-2s\left[ (r+ia\cos
\theta )-\frac{M(r^2-a^2)}\Delta \right] \partial_t  \nonumber \\
&&+\,\Delta^{-s}\partial_r\left( \Delta^{s+1}\partial_r\right) +
\frac 1{\sin \theta }\partial_\theta \left( \sin \theta \partial_\theta
\right) +\left( \frac 1{\sin^2\theta }-\frac{a^2}\Delta \right) 
\partial_{\varphi \varphi } \nonumber \\
&&+\,2s\left[ \frac{a(r-M)}\Delta +\frac{i\cos \theta }{\sin^2\theta }
\right] \partial_\varphi -\left( s^2\cot^2\theta -s\right) \Biggr\}\Psi
=4\pi \Sigma T \;,  \label{master}
\end{eqnarray}
where $M$ is the mass of the black hole, $a$ its angular momentum per
unit mass, $s$ the spin of the perturbation, $\Sigma \equiv
r^2+a^2\cos^2\theta$, and $\Delta
\equiv  r^2-2Mr+a^2$.
The source term $T$ is built up from the energy-momentum
tensor~\cite{Teukolsky73}.
Gravitational perturbations, corresponding
to $s=\pm2$, are compactly described in terms of contractions of the
Weyl tensor with a null tetrad.  The components of the tetrad (also
given in Ref.\ \cite{Teukolsky73}) are appropriately chosen along the
repeated principal null directions of the background spacetime [see
Eq.\ (\ref{tetrad}) below].  The resulting (infinitesimal) coordinate and
tetrad invariant components of the Weyl curvature are given by
\begin{equation}\label{psi}
\Psi=\left\{ 
\begin{array}{ll}
\rho_K^{-4}\psi_4\equiv -\rho_K^{-4}C_{\mathrm{n\overline mn\overline m}} & {\rm for}~~s=-2
\\ 
\psi_0\equiv -C_{\mathrm{lmlm}} & {\rm for}~~s=+2~
\end{array}
\right. , 
\end{equation}
where an overbar means complex conjugation and $\rho_K$ is given in Eq.\
(\ref{spincoeff}) below. 
Asymptotically, the leading behavior of the field $\Psi$
represents either the outgoing radiative part of the perturbed Weyl tensor, 
($s=-2$), or the ingoing radiative part, ($s=+2$).

The components of the Boyer--Lindquist null tetrad for the Kerr
background are given by
\begin{subequations}\label{tetrad}
\begin{eqnarray}
   (\mathrm{l_K}^{\alpha}) &=& \left( \frac{r^2+a^2}{\Delta},1,0,\frac{a}{\Delta} 
                  \right), \\
   (\mathrm{n_K}^{\alpha}) &=& \frac{1}{2\;\!(r^2+a^2\cos^2\theta)} \,
                  \left( r^2+a^2,-\Delta,0,a \right), \\
   (\mathrm{m_K}^{\alpha}) &=& \frac{1}{\sqrt{2}(r+ia\cos\theta)}\,
                  \left( ia\sin\theta,0,1,i/\sin\theta \right).\quad\quad 
\end{eqnarray}
\end{subequations}
One can also define directional derivatives as
\begin{equation}
\delta=\mathrm{m_K}^\mu\partial_\mu~;~~
\hat\Delta=\mathrm{n_K}^\mu\partial_\mu~;~~
\hat{D}=\mathrm{l_K}^\mu\partial_\mu~.
\end{equation}

With the above choice of the tetrad the non-vanishing spin coefficients are
(where an overbar stands for complex conjugation)
\begin{eqnarray}\label{spincoeff}
\rho_K&=&-\frac{1}{(r-ia\cos\theta)},\quad
\beta_K=-\overline{\rho}_K\frac{\cot\theta}{2\sqrt{2}},\nonumber\\
\pi_K&=&ia\rho_K^2\frac{\sin\theta}{\sqrt{2}},\quad
\tau_K=-ia\rho_K\overline{\rho}_K\frac{\sin\theta}{\sqrt{2}},\nonumber\\
\mu_K&=&\rho_K^2\overline{\rho}_K\ \frac{\Delta}{2},\quad
\alpha_K=\pi_K-\overline{\beta}_K,\nonumber\\
\gamma&=&\mu+\rho_K\overline{\rho}_K\frac{(r-M)}{2},
\end{eqnarray}
and the only non-vanishing Weyl scalar in the background is
\begin{equation}\label{psi2} 
\psi_2=M\rho_K^3.
\end{equation}

As we mentioned above the Weyl scalars $\psi_4$ and $\psi_0$ allow a
direct computation of the radiation escaping to infinity
\cite{Campanelli99} and going down the horizon. The time-domain
formulation is particularly well suited for interfacing with full
numerical relativity techniques
\cite{Baker00b,Baker:2001nu,Baker:2001sf,Baker:2002qf,Baker:2004wv}.
There are, besides, other physical phenomena of interest such as the
self-force on a particle orbiting the hole \cite{Lousto99b}, studies
of the horizon structure, and second order perturbations
\cite{Campanelli99}, that require the computation of the metric
perturbations.

Starting from the pioneering work of Chzranowski
\cite{Chrzanowski:1975wv} there is a series of papers
\cite{Kegeles:1979an,1979RSPSA.369...67W,1979RSPSA.367..527S,Ori:2002uv}
dealing with the problem of metric reconstruction by the introduction
of a potential that satisfies the Teukolsky equation, being neither
the $\psi_4$ or $\psi_0$ describing to the physical situation under
study.  The problem of relating the potentials introduced to describe
the {\it radiation} gauges and the physical $\psi_4$ and $\psi_0$ has
been recently studied in Ref.~\cite{Lousto:2002em}. There, the results
are explicitly given for vacuum metric perturbations on the
Schwarzschild, i.e. nonrotating, black hole background.

Given the difficulties in obtaining by these method the explicit
metric expression for perturbations around a Kerr, i.e. rotating black
hole, background, we present here an alternative approach. In this
paper we give explicit formulae for the metric reconstruction in the
Regge-Wheeler gauge, still for a nonrotating background, but allowing
for source terms, bearing in mind, for instance, the applications to
the radiation reaction problem. In the next section we will
give explicitly the form $\psi_4$ and $\psi_0$ take in terms of metric
perturbation, in the Regge-Wheeler gauge, making explicit use to the
multipole decomposition of the metric. In order to invert these
expressions, in Sec.~\ref{RWG} we introduce the symmetric and
antisymmetric components of the Weyl scalar under the discrete
parity transformation. With the help of the field equations of
General Relativity in the Regge-Wheeler gauge (reviewed in
\ref{HEEquations}) we succeed in expressing the metric perturbations in
terms of $\psi_4$ and $\psi_0$, including matter terms. In the final
section of the paper we describe how to generalize the first few of
these steps to the Kerr background case, and speculate about the
completion of this program.

\section{Weyl Scalars}\label{Weyl}

The first step in explicitly constructing the metric perturbations is
actually computing the inverse relation, that makes use of the definition
of the Weyl scalars (\ref{psi}) in terms of the Weyl tensor.
Chrzanowski \cite{Chrzanowski:1975wv}, made this computation explicitly
relating the perturbed Weyl scalars to the metric perturbations

\begin{eqnarray}\label{psi4}
\psi_4 &=&\frac 12\Biggr\{(\overline \delta +3\alpha +
\overline \beta- \overline \tau)(\overline \delta +2\alpha +2\overline \beta -\overline \tau )h_{\mathrm{nn}}\nonumber\\
&&+(\hat \triangle +\overline \mu +3\gamma -\overline \gamma )
(\hat \triangle +\overline \mu +2\gamma -2\overline \gamma ) h_{\mathrm{\overline m\overline m}}\nonumber\\
&&-\left[ (\hat \triangle +\overline \mu +3\gamma -\overline\gamma )
(\overline\delta-2\overline \tau +2\alpha )\right.\nonumber\\
&&+\left.(\overline\delta +3\alpha +\overline\beta -\overline\tau )
(\hat \triangle +2\overline \mu +2\gamma )\right] h_{(\mathrm{n\overline m})}\Biggr\},
\end{eqnarray}


and

\begin{eqnarray}\label{psi0}
\psi_0 &=&\frac 12\Biggr\{( \delta -\overline \alpha-3\beta+\overline\pi )
( \delta -2\overline\alpha -2\beta +\overline \pi )h_{\mathrm{ll}}\nonumber\\
&&+(\hat D-\overline \rho-3\epsilon+\overline\epsilon )
(\hat D-\overline \rho-2\epsilon+2\overline\epsilon )h_{\mathrm{mm}}  \nonumber
\\
&&-\left[ (\hat D-\overline \rho -3\epsilon+\overline\epsilon)
(\delta+2\overline\pi-2\beta )\right.\nonumber\\
&&+\left.( \delta -\overline \alpha -3\beta+\overline\pi )
(\hat D-2\overline\rho-2\epsilon)\right] h_{(\mathrm{lm})}\Biggr\},
\end{eqnarray}
where $h_{\mathrm{nn}}=\mathrm{n}^\mu \mathrm{n}^\nu h_{\mu \nu }$, $h_{\mathrm{lm}}=\mathrm{l}^\mu \mathrm{m}^\nu h_{\mu \nu }$,
etc.

Many simplifications are possible in the analysis when the background has
spherical symmetry. In the Schwarzschild black hole case expressions \ (\ref
{psi4})-(\ref{psi0}) reduce to 
\begin{eqnarray}   \label{psis4}
\psi _4&=&\frac 1{16}\Biggr\{\frac 1{r^2}\left( \partial _\theta
-\cot \theta -\frac i{\sin \theta }\partial _\varphi \right)
\left(\partial _\theta -\frac i{\sin \theta }\partial _\varphi
\right)\nonumber\\
&&\times\left[h_{tt}-2h_{rt}f+h_{rr}f^2\right]   \nonumber \\
&&+\left( \partial _t-f\partial _r+f^{\prime }-\frac{2f}{r}\right)
\left( \partial _t-f\partial _r\right)\nonumber\\
&&\times\frac 1{r^2}\left[ h_{\theta
\theta }-\frac{h_{\varphi \varphi }}{\sin ^2\theta }-2i\frac{h_{\theta
\varphi }}{\sin \theta }\right]\nonumber\\
&&-\frac 2{r^2}\left( \partial _t-f\partial _r+f^{\prime}
\right)\left( \partial _\theta-\cot \theta -\frac i{\sin \theta }
\partial_\varphi \right)  \nonumber \\
&&\times\left[
h_{t\theta }-i\frac{h_{t\varphi }}{\sin \theta }-f\left( h_{r\theta }
-i\frac{h_{r\varphi }}{\sin \theta }\right) \right] \Biggr\},
\end{eqnarray}
and 
\begin{eqnarray}\label{psis0}
\psi _0 &=&\frac 1{4}\Biggr\{ \frac 1{r^2}\left( \partial _\theta -\cot \theta +
\frac i{\sin \theta }\partial _\varphi \right)
\left(\partial _\theta +\frac i{\sin \theta }\partial _\varphi\right)\nonumber\\
&&\times\left[h_{tt}f^{-2}+2h_{rt}f^{-1}+h_{rr}\right]   \nonumber \\
&&+\left( f^{-1}\partial _t+\partial _r+\frac 2r\right)
\left( f^{-1}\partial _t+\partial _r\right)\nonumber \\
&&\times\frac 1{r^2}\left[h_{\theta \theta }-\frac{h_{\varphi \varphi }}
{\sin ^2\theta }+2i\frac{h_{\theta \varphi }}{\sin \theta }\right]   \nonumber\\
&&+\frac 2{r^2}\left( f^{-1}\partial _t+\partial _r
\right)\left( \partial _\theta-\cot \theta +\frac i{\sin \theta }
\partial_\varphi \right)  \nonumber \\
&&\times\left[ f^{-1}
\left(h_{t\theta }+i\frac{h_{t\varphi }}{\sin \theta }\right) +h_{r\theta }+i\frac{
h_{r\varphi }}{\sin \theta }\right]\Biggr\},
\end{eqnarray}
where $f=1-2M/r$ and $f^{\prime }=2M/r^2$ .

The imposition of spherical symmetry also carries the following
computational 
advantage: the multipole
decomposition of the metric perturbations in terms of spin-weighted
harmonics $_{-2}Y_{\ell m}(\theta)$ can be
performed\cite{Regge57,Moncrief74}, and even and odd parity
perturbations decouple so they can be considered independently. Below
we shall decompose all metric perturbations in multipoles with
index $\ell m$ (not to be confused with the tetrad
vectors).

From Eq. (\ref{psis4}) and (\ref{psis0}) in the Regge-Wheeler gauge
$(h_1^{\ell m}=h_0^{\ell m}=G^{\ell m}=0={}^{(odd)}h_2^{\ell m})$, we
get

\begin{eqnarray}\label{RWSpsi4}
\psi_4&\dot=&\sum_{\ell m} \psi_4^{\ell m}\;_{-2}Y_{\ell m}=
  -\sum_{\ell m}\sqrt{\frac{(\ell-2)!}{(\ell+2)!}}\nonumber\\
&&\times\left\{-\frac{f}{16r^2}\left(H_0^{\ell m}-2H_1^{\ell m}
+H_2^{\ell m}\right)\right.\nonumber\\
&&\left.-\frac i{8r^2}\left[ \partial _t-f \partial _r+f'
\right]\left({}^{(odd)}h_0^{\ell m}-f\;{}^{(odd)}h_1^{\ell m}\right)\right\}\;_{-2}Y_{\ell m},
\end{eqnarray}

and

\begin{eqnarray}\label{RWSpsi0}
\psi_0 &\dot=&\sum_{\ell m} \psi_0^{\ell m}\;_{+2}Y_{\ell m}=
-\sum_{\ell m}\sqrt{\frac{(\ell-2)!}{(\ell+2)!}}\nonumber\\
&&\times\left\{-\frac{1}{4fr^2}\left(H_0^{\ell m}+2H_1^{\ell m}+H_2^{\ell m}\right)\right.\nonumber\\
&&\left.-\frac i{2f^2r^2}\left[ \partial _t+f \partial _r-f'
\right]\left({}^{(odd)}h_0^{\ell m}+f\;{}^{(odd)}h_1^{\ell m}\right)\right\}\;_{+2}Y_{\ell m}.
\end{eqnarray}

These represent our basic equations (real and imaginary parts) that we
will use in the next section to express the metric perturbations in
terms of $\psi_4$ and $\psi_0$.

\section{Explicit solution in the Regge-Wheeler gauge}\label{RWG}

Two key elements are introduced here in order to complete the
inversion of metric coefficients from Eqs. \ref{RWSpsi4} and
\ref{RWSpsi0}. The first is the decomposition of the Weyl fields into
its symmetric and antisymmetric parts with respect to the 
discrete parity transformation. This allows to separate the even and odd
parity perturbations from the multipole decomposed $\psi_4$, $\psi_0$,
and metric perturbations. For the even parity case this allows to
obtain directly two of the four metric coefficients. In order to
obtain the other two, we have to resource to the General Relativity
field equations, which represent the other key element in the
inversion process.  For the odd parity case, one ends up with first
order differential relations that can be brought to explicit
integrals (previous simplification by making us of the odd parity
field equations).

\subsection{Even parity}

Let us define the symmetric and antisymmetric Weyl scalar fields as
\cite{Lousto:2002em}
\begin{eqnarray}\label{pm}
\psi^\pm=\frac12\left[\psi^{\ell,m}\pm(-)^m\overline{\psi}^{\ell,-m}\right],
\end{eqnarray}
where for notational simplicity we dropped the $\ell,m$ indexes. 
Thus, given the symmetric nature of the even parity metric perturbations,
Eqs.~(\ref{RWSpsi4}) and (\ref{RWSpsi0}) take the form
\begin{eqnarray}\label{RWSpsi4+}
\psi_4^+&=&\frac{f}{16r^2}
\sqrt{\frac{(\ell-2)!}{(\ell+2)!}}\left(H_0^{\ell m}-2H_1^{\ell m}
+H_2^{\ell m}\right),
\end{eqnarray}
and
\begin{eqnarray}\label{RWSpsi0+}
\psi_0^+&=&\frac{1}{4fr^2}
\sqrt{\frac{(\ell-2)!}{(\ell+2)!}}
\left(H_0^{\ell m}+2H_1^{\ell m}+H_2^{\ell m}\right).
\end{eqnarray}

From Eqs.\ (\ref{RWSpsi4+}) and \ (\ref{RWSpsi0+}) we can obtain the
$\ell m$ components of the metric perturbations as follows

\begin{eqnarray}\label{H1psi}
H_1^{\ell m}(r,t)=
-\frac{4r^2}{f}\sqrt{\frac{(\ell+2)!}{(\ell-2)!}}
\left[\psi_4^{+}-\frac{f^2}{4}\psi_0^{+}\right] 
\end{eqnarray}
and 
\begin{eqnarray}
H_0^{\ell m}(r,t)&+&H_2^{\ell m}(r,t)=
\frac{8r^2}{f}\sqrt{\frac{(\ell+2)!}{(\ell-2)!}}
\left[\psi_4^{+}+\frac{f^2}{4}\psi_0^{+}\right] 
\end{eqnarray}

We now bring into the play
the Hilbert-Einstein's equations in the Regge-Wheeler gauge,
Eq.~(\ref{C7g})
give,
\begin{equation}\label{RWEvenEinstein}
H_0^{\ell m}(r,t)-H_2^{\ell m}(r,t)=
\frac{16\pi r^2}{\sqrt{2\lambda(\lambda+1)}}F^{\ell m}~,  
\end{equation}
where $F_{\ell m}$is a source term given in Table III of
Ref.\cite{Zerilli70} (See also \ref{Recon}.)  This allows to
find the metric perturbations for $H_0^{\ell m},$ $H_1^{\ell m},$ and
$H_2^{\ell m}$. The last metric coefficient has to be found by use of
the Hilbert-Einstein equations [$K^{\ell m}$ and $H_0^{\ell
m}-H_2^{\ell m}$ give a measure of the trace of the even parity
sector in the Regge-Wheeler gauge, so it does not appear in the Weyl
scalars since the Weyl tensor is traceless.]

Using Eq.~(\ref{C7e})
we can solve for $\partial_r K^{\ell m}$ and then replace
it in Eq.\ (\ref{C7a}) to find $K^{\ell m}$ in terms of the other
even parity metric coefficients and source terms 
\begin{eqnarray}\label{K}
&&K(r,t)^{\ell m}=\nonumber\\
&&2\,{\frac {\left (r-M\right ){\frac {\partial }{\partial r}}{H_0^{\ell m}}(r
,t)}{\lambda}}+{\frac {\left (r-2\,M\right )r{\frac {\partial ^{2}}{
\partial {r}^{2}}}{H_0^{\ell m}}(r,t)}{\lambda}}\nonumber\\
&&-{\frac {{r}^{2}{\frac {
\partial ^{2}}{\partial r\partial t}}{H_1^{\ell m}}(r,t)}{\lambda}}+{\frac {
M{\frac {\partial }{\partial r}}{H_2^{\ell
m}}(r,t)}{\lambda}}\nonumber\\ &&-{\frac {\left
(2\,{r}^{2}-8\,rM+9\,{M}^{ 2}\right){H_0^{\ell
m}}(r,t)}{r\lambda\,\left (r-2\,M\right )}}\nonumber\\ &&+{\frac
{\left (-{r}^{2}\lambda+2\,rM\lambda+3\,{M}^{2}-2\,rM\right
){H_2^{\ell m}}(r ,t)}{r\lambda\,\left(r-2\,M\right)}}\nonumber\\
&&+{\frac {r\left (-3\,r+7\,M
\right){\frac {\partial }{\partial t}}{H_1}(r,t)}{\lambda\,\left(r-2\,M\right)}}
+8\,{\frac {\left (r-2\,M\right )\pi \,{r}^{2}{\frac {\partial }{\partial r}}B^{\ell m}(r,t
)}{\sqrt {\lambda+1}\lambda}}\nonumber\\
&&-8\,{\frac {\left (7\,M-4\,r\right )\pi \,rB^{\ell m}(r,t)}{
\sqrt {\lambda+1}\lambda}}-{\frac {A_0^{\ell m}(r,t){r}^{3}}{\lambda\,
\left (r-2\,M\right )}}.
\end{eqnarray}

This form of the metric coefficient $K^{\ell m}$, involves second
derivatives of the Weyl scalars. One can consider an alternative
integral form (on the hypersurface $t=constant$) derived from
Eq.~(\ref{C7e})

\begin{eqnarray}\label{Kbis}
K^{\ell m}&=&H_0^{\ell m}+\int_{2M}^{r}\frac{dr}{1-\frac{2M}{r}}\left[
-\frac{\partial H_1^{\ell m}}{\partial t}+\frac{2M}{r^2}H_0^{\ell
m}\right. \\
&&\left.
-16\pi\left(r-2M\right)\frac{F^{\ell
m}}{\sqrt{2\lambda(\lambda+1)}}
-\frac{8\pi(r-2M)}{\sqrt{\lambda+1}}B^{\ell m}\right]\nonumber
\end{eqnarray}

It is worth mentioning here that if the source is modeled as a
particle (represented by a Dirac's Delta) the above metric
coefficients are continuous ($C^0$) for headon
collisions\cite{Lousto99b} at the location of the particle. However,
for more general orbits they do not all appear to be continuous but
some of them behave as a Dirac's Delta. For instance, one can see that
from expression (\ref{RWEvenEinstein}). ${F^{\ell m}}$ is proportional
to a Dirac's Delta, as given in Table~\ref{Tmunu}; hence, at least
$H_2^{\ell m}$ or $H_2^{\ell m}$ have to behave as $\delta[r-R(t)]$.

\subsection{Odd parity}

From Eq. (\ref{RWSpsi4}) and (\ref{RWSpsi0}), given the antisymmetric
behaviour of the odd parity metric coefficients, we get

\begin{eqnarray}\label{RWpsi4}
\psi_4^-&=&\frac i{8r^2}\sqrt{\frac{(\ell-2)!}{(\ell+2)!}}
\left[\partial_t-f\partial_r+f'
\right]\left({}^{(odd)}h_0^{\ell m}-f\;{}^{(odd)}h_1^{\ell m}\right),
\end{eqnarray}
and
\begin{eqnarray}\label{RWpsi0}
\psi_0^-&=&\frac {i}{2f^2r^2}\sqrt{\frac{(\ell-2)!}{(\ell+2)!}}
\left[\partial_t+f\partial_r-f'
\right]\left({}^{(odd)}h_0^{\ell m}+f\;{}^{(odd)}h_1^{\ell m}\right).
\end{eqnarray}

A linear combination of these previous equations produces

\begin{eqnarray}\label{psi-+}
&&\psi_4^-+\frac{f^2}{4}\psi_0^-=\frac i{4r^2}\sqrt{\frac{(\ell-2)!}{(\ell+2)!}}
\left[\partial_t{}^{(odd)}h_0^{\ell m}+
\left(f\partial_r-f'
\right)\left(f\;{}^{(odd)}h_1^{\ell m}\right)\right],
\end{eqnarray}
and
\begin{eqnarray}\label{psi--}
&&\psi_4^--\frac{f^2}{4}\psi_0^-=\frac {-i}{4r^2}\sqrt{\frac{(\ell-2)!}{(\ell+2)!}}\left[f\partial_t{}^{(odd)}h_1^{\ell m}+
\left(f\partial_r-f'
\right)\left(\;{}^{(odd)}h_0^{\ell m}\right)\right].
\end{eqnarray}

From Eq.~(\ref{thetaphiodd}) we can substitute $\partial_t h_0^{\ell m}$ into Eq.~(\ref{psi-+})
leading to the equation

\begin{eqnarray}\label{h1diff}
&&\partial_r h_1^{\ell m}+\left(\frac{f'}{2f}
\right)h_1^{\ell m} = S_1^{\ell m}(r,t)\dot=\nonumber\\
&&\frac{-2ir^2}{f^2}\sqrt{\frac{(\ell+2)!}{(\ell-2)!}}
\left(\psi_4^-+\frac{f^2}{4}\psi_0^-\right)-
\frac{2\pi\,ir^2D_{\ell m}}{f\sqrt{\lambda(\lambda+1)}},
\end{eqnarray}

which integrated produces

\begin{eqnarray}\label{h1}
h_1^{\ell m}=\frac{1}{\sqrt{1-\frac{2M}{r}}}\left\{
\int_{2M}^r\,S_1^{\ell m}(r',t)\,\sqrt{1-\frac{2M}{r'}}\,dr'\,+\,C_1^{\ell m}(t)
\right\},\\
\end{eqnarray}
where $C_1^{\ell m}$ is an integration constant, that in vacuum and with
vanishing $\psi_0$ and $\psi_4$ can be taken to vanish \cite{Wald:1973}.

Knowing now the form of $h_1^{\ell m}$ we can use Eq~(\ref{psi--}) to find a differential
equation for $h_0^{\ell m}$ 

\begin{eqnarray}\label{h0diff}
&&\partial_r h_0^{\ell m}-\left(\frac{f'}{f}
\right)h_0^{\ell m} = S_0^{\ell m}(r,t)\dot=\frac{4ir^2}{f}\sqrt{\frac{(\ell+2)!}{(\ell-2)!}}
\left(\psi_4^--\frac{f^2}{4}\psi_0^-\right)-\partial_t\,h_1^{\ell m},
\nonumber\\
\end{eqnarray}

with solution

\begin{eqnarray}\label{h0}
h_0^{\ell m}=\left(1-\frac{2M}{r}\right)\left\{
\int_{2M}^r\,\frac{S_0^{\ell m}(r',t)}{1-\frac{2M}{r'}}\,dr'\,+\,C_0^{\ell m}(t)
\right\}.\\
\end{eqnarray}
Again, $C_0^{\ell m}$ is an integration constant, that in vacuum and
with vanishing $\psi_0$ and $\psi_4$ can be taken to vanish.

This essentially completes the work of expressing the metric
perturbations in terms of the computed $\psi_0$ and $\psi_4$ expressed
in the time domain. For the even parity case it contains second
derivatives of the Weyl scalars (unlike the corresponding expressions
for the radiation gauge
\cite{Lousto:2002em}.)  For the odd parity case, solutions (\ref{h0})
and (\ref{h1}) are written in an integral form.

A last observation applies here, since the spin weight of the Weyl
scalars are $s=\pm2$ they do not contain multipole modes $\ell=0$ and
$\ell=1$, hence we need to give them by directly solving the field
equations for the metric coefficients. The Regge-Wheeler gauge do not
completely allow to determine them, because there are one extra degree
of freedom for $\ell=1$ and two degrees of freedom for
$\ell=0$. Zerilli
\cite{Zerilli70} has made choices to fix this extra freedom that
allowed him to solve analytically for the metric coefficients. In
Ref. \cite{BL02a} a different choice was made to make those
coefficients continuous in the headon collision of extreme mass black
holes. Finally in Ref.~\cite{Detweiler:2003ci} the metric
coefficients for $\ell=0,1$ have been found in the harmonic gauge, for
particles in circular orbits.


\section{Discussion of Kerr perturbations}\label{discussion}

A possible generalization of the Regge-Wheeler gauge conditions
for spherically symmetric backgrounds, but where perturbations are
not decomposed into multipoles is \cite{Barack:2001ph}
\begin{eqnarray}
(\sin\theta)^2 h_{\theta\theta}-h_{\phi\phi}&=&0,\label{RWg1}\\
h_{\theta\phi}&=&0,\label{RWg2}\\
\sin\theta\;\partial_\theta(\sin\theta h_{t\theta})+\partial_\phi h_{t\phi}&=&0,\label{RWg3}\\
\sin\theta\;\partial_\theta(\sin\theta h_{r\theta})+\partial_\phi h_{r\phi}&=&0.\label{RWg4}
\end{eqnarray}

The first equation above leads to the condition $G^{\ell m}=0$. The
second gives then ${}^{(odd)}h_2^{\ell m}=0$. The other two
differential conditions are chosen such that they lead to
${}^{(even)}h_0^{\ell m}=0={}^{(even)}h_1^{\ell m}$, but allow
${}^{(odd)}h_0^{\ell m}\not=0$ and ${}^{(odd)}h_1^{\ell m}\not=0$
be unconstrained.

Now we will consider the generalization of the Regge-Wheeler gauge in
the Newman-Penrose formalism.  In this formalism, the
first two Regge-Wheeler conditions, Eqs.~(\ref{RWg1}) and (\ref{RWg2}),
have a simple generalization
\begin{eqnarray}\label{rw1}
h_{\mathrm{mm}}=\mathrm{m}^\mu \mathrm{m}^\nu h_{\mu \nu }=0.
\end{eqnarray}
Note that requiring that the real and imaginary parts vanish
contains both conditions. Obviously, 
\begin{eqnarray}\label{rw2}
h_{\mathrm{\overline{m}\overline{m}}}=\mathrm{\overline{m}}^\mu \overline{\mathrm{m}}^\nu h_{\mu \nu }=0,
\end{eqnarray}
also holds. Note that conditions (\ref{rw1}) and (\ref{rw2}) are
invariant under type III (spin-boosts) transformations of the
background tetrad
\begin{eqnarray}\label{typeIII}
\mathrm{l}\to A^2\mathrm{l},\quad\quad \mathrm{n}\to A^{-2}\mathrm{n},
\quad\quad \mathrm{m}\to e^{2i\Theta}\mathrm{m},\quad\quad
\mathrm{\bar{m}}\to e^{-2i\Theta}\mathrm{\bar{m}}.
\end{eqnarray}

This is an important feature, since the Kinnersley
choice of the tetrad, with the spin coefficient $\epsilon=0$ is just a
simple, but arbitrary way of fixing the spin-boost freedom. In
contrast the convenient choice of the $l$ and $n$ tetrad vectors along
the repeated principal null directions of the Kerr background allows
to single out wave equations for the perturbations of $\psi_4$ and
$\psi_0$.

To generalize the differential conditions (\ref{RWg3}) and
(\ref{RWg4}) one can resort to the type III transformation properties
of the $\delta$ and $\bar\delta$, as well as spin coefficient
operators in the Kerr background acting on the metric coefficients
$h_{(\mathrm{lm})}$ and $h_{(\mathrm{n\overline m})}$. The objects
\begin{eqnarray}\label{transf1}
(\delta-2\bar\alpha)h_{(\mathrm{l\bar{m}})}\to
A^2(\delta-2\bar\alpha)h_{(\mathrm{l\bar{m}})},
\end{eqnarray}
and
\begin{eqnarray}\label{transf2}
(\bar\delta+2\bar\beta)h_{(\mathrm{nm})}\to
A^{-2}(\bar\delta+2\bar\beta)h_{(\mathrm{nm})},
\end{eqnarray}
transform as objects of spin- $0$ and boost weight $+1$ and $-1$
respectively under type III transformations of the background tetrad
(\ref{typeIII}).

In order to reproduce the differential conditions (\ref{RWg3}) and
(\ref{RWg4}) in the Schwarzschild limit, one can then require
\begin{eqnarray}\label{rw3}
\Re\left[
(\delta-2\bar\alpha+a\tau-b\bar\pi)h_{(\mathrm{l\bar{m}})}
\right]=0,
\end{eqnarray}
and 
\begin{eqnarray}\label{rw4}
\Re\left[
(\bar\delta+2\bar\beta+c\bar\tau-d\pi)h_{(\mathrm{nm})}
\right]=0,
\end{eqnarray}
Where $\Re$ is the real part and where following \cite{WP} we
introduced additional terms containing spin coefficients with spin
$\pm1$ respectively and boost $0$ multiplied by constants $a,b,c,d$ to
allow for a more general choice of the gauge. These constants can be
readily chosen to facilitate the metric reconstruction or, in other
contexts, to impose further symmetries or facilitate the numerical
integration of General Relativity field equations, etc. It also
stresses the ambiguities in generalizing the Regge-Wheeler gauge on
the Kerr background.

Note also that conditions (\ref{rw3}) and (\ref{rw4}) are invariant
under type III transformations as well.  A crucial role in achieving
that was played by the spin- $0$ transformation properties of the
constructed object, allowing invariance of its Real part.

Independently, we can try to proceed along the lines of the previous
sections with now a simple mode decomposition of the metric
coefficients. For instance
\begin{eqnarray}\label{skerr}
h_{(\mathrm{lm})}^\pm=\frac12\left[h^m_{(\mathrm{lm})}
\pm (-)^m h^{-m}_{(\mathrm{l\overline{m}})}\right]
\end{eqnarray}
where we decomposed
\begin{eqnarray}\label{mkerr}
h_{(\mathrm{lm})}(t,r,\theta,\varphi)=\sum_{m}e^{im\varphi}
h^m_{(\mathrm{lm})}(t,r,\theta)
\end{eqnarray}

We can now replace this decomposition directly into Eqs.~(\ref{psi4})
and (\ref{psi0}) for $\psi_4$ and $\psi_0$ or the following more
convenient form making use of Ref.~\cite{Teukolsky73}, Eq.~(2.11),
where we have the following identity
\begin{eqnarray}
\left[\hat{D}-(p+1)\epsilon+\overline{\epsilon}+q\rho-\overline\rho\right]
\left(\delta-p\beta+q\tau\right)\nonumber\\
=\left[\delta-(p+1)\beta-\overline{\alpha}+q\tau+\overline\pi\right]
\left(\hat{D}-p\epsilon+q\rho\right),
\end{eqnarray}
and the identity derived from it exchanging tetrads $\mathrm{l\to n}$
and $\mathrm{m\to \overline{m}}$.
\begin{eqnarray}
\left[\hat\Delta+(p+1)\gamma-\overline{\gamma}-q\mu+\overline\mu\right]
\left(\overline\delta+p\alpha-q\pi\right)\nonumber\\
=\left[\overline\delta+(p+1)\alpha+\overline{\beta}-q\pi-\overline\tau\right]
\left(\hat\Delta+p\gamma-q\mu\right).
\end{eqnarray}

Using $p=2$ and $q=0$ in the above identities allows us to rewrite
Eqs.~(\ref{psi4}) and (\ref{psi0}) as
\begin{eqnarray}\label{psi4b}
\psi_4 &=&\frac 12\Biggr\{(\overline \delta +3\alpha +
\overline \beta- \overline \tau)(\overline \delta +2\alpha +2\overline \beta -\overline \tau )h_{\mathrm{nn}}\nonumber\\
&&+(\hat \triangle +\overline \mu +3\gamma -\overline \gamma )
(\hat \triangle +\overline \mu +2\gamma -2\overline \gamma ) h_{\mathrm{\overline m\overline m}}\nonumber\\
&&-\left[2(\hat \triangle +\overline \mu +3\gamma -\overline\gamma )
(\overline\delta-\overline \tau +2\alpha )\right.\nonumber\\
&&+2\left.(\overline\delta +3\alpha +\overline\beta -\overline\tau )
(\overline\mu)\right] h_{(\mathrm{n\overline m})}\Biggr\},
\end{eqnarray}
and
\begin{eqnarray}\label{psi0b}
\psi_0 &=&\frac 12\Biggr\{( \delta -\overline \alpha-3\beta+\overline\pi )
( \delta -2\overline\alpha -2\beta +\overline \pi )h_{\mathrm{ll}}\nonumber\\
&&+(\hat D-\overline \rho-3\epsilon+\overline\epsilon )
(\hat D-\overline \rho-2\epsilon+2\overline\epsilon )h_{\mathrm{mm}}  \nonumber
\\
&&-\left[ 2(\hat D-\overline \rho -3\epsilon+\overline\epsilon)
(\delta+\overline\pi-2\beta )\right.\nonumber\\
&&-2\left.( \delta -\overline \alpha -3\beta+\overline\pi )
(\overline\rho)\right] h_{(\mathrm{lm})}\Biggr\}.
\end{eqnarray}


A choice of the symmetric tetrad \ref{symtetrad} will further simplify
the appearance of the equations. At this point we impose our gauge
condition on $h^+_{(\mathrm{lm})}$ and $h^+_{(\mathrm{n\overline
m})}$.  Then, paralleling the work done in the nonrotating limit, we
could make further progress by writing explicitly the Newman-Penrose
equations in terms of metric perturbations, for finally using these
equations to obtain decoupled expressions for some metric
coefficients. The completion of this program remains an open issue and
goes beyond the scope of this paper. We leave this for future research.

\ack
The author specially thanks L.~Barack, L.~Price and B.~Whiting for
useful discussions during the development of this manuscript.
C.O.L. gratefully acknowledges the support of the NASA Center for
Gravitational Wave Astronomy at The University of Texas at Brownsville
(NAG5-13396), and from NSF grants PHY-0140326 and PHY-0354867.


\appendix
\section{Hilbert-Einstein equations in the Regge-Wheeler gauge}\label{HEEquations}

Because of some misprints in the original Zerilli's paper
\cite{Zerilli70} we reproduce here the relevant equations for
our discussions (See also Ref.~\cite{1978pans.proc.....G,Sago:2002fe}).

The metric perturbations on a Schwarzschild background
\begin{eqnarray}
ds^2=&-&(1-\frac{2M}{r})\,dt^2+(1-\frac{2M}{r})^{-1}\,dr^2
+r^2\,\left(d\theta^2+\sin\theta\,d\varphi^2\right),
\end{eqnarray}
can be decomposed into spherical harmonics\cite{Zerilli70}
\renewcommand{\arraystretch}{1.5}
\begin{equation}\label{evenperts}
h_{\mu\nu}^{\rm{even}}=
\begin{vmatrix}
(1-\frac{2M}{r})H_0^{\ell m}(t,r) & H_1^{\ell m}(t,r) & 
h_0^{\ell m}(t,r)\,\frac{\partial}{\partial\theta} & 
h_0^{\ell m}(t,r)\, \frac{\partial}{\partial\varphi} \\
H_1^{\ell m}(t,r) & \frac{ H_2^{\ell m}(t,r)}{(1-\frac{2M}{r})} &
h_1^{\ell m}(t,r)\,\frac{\partial}{\partial\theta} & 
h_1^{\ell m}(t,r)\,\frac{\partial}{\partial\varphi} \\
h_0^{\ell m}(t,r)\,\frac{\partial}{\partial\theta} & 
h_1^{\ell m}(t,r)\,\frac{\partial}{\partial\theta} & 
r^2\left[K^{\ell m}\,+\,G^{\ell m}\,\frac{\partial^2}{\partial\theta^2}\right] &
r^2\,G^{\ell m}(t,r)\,\frac{\hat{X}_{\ell m}}{2}\\
h_0^{\ell m}(t,r)\, \frac{\partial}{\partial\varphi} & 
h_1^{\ell m}(t,r)\,\frac{\partial}{\partial\varphi} &
r^2\,G^{\ell m}(t,r)\,\frac{\hat{X}_{\ell m}}{2} &
r^2\left[K^{\ell m}\sin^2\theta\,+\,G^{\ell m}\,\hat{Z}_{\ell m}\right]
\end{vmatrix}
Y_{\ell m}(\theta,\varphi)
\end{equation}
for the {\it even} parity modes.

And
\begin{equation}\label{oddperts}
h_{\mu\nu}^{\rm{odd}}=
\begin{vmatrix}
0 &
0 &
-h_0^{\ell m}(t,r)\,\frac{\partial}{\sin\theta\partial\varphi} &
h_0^{\ell m}(t,r)\,\frac{\sin\theta\partial}{\partial\theta} \\
0 &
0 &
-h_1^{\ell m}(t,r)\,\frac{\partial}{\sin\theta\partial\varphi} &
h_1^{\ell m}(t,r)\,\frac{\sin\theta\partial}{\partial\theta} \\
-h_0^{\ell m}(t,r)\,\frac{\partial}{\sin\theta\partial\varphi} &
-h_1^{\ell m}(t,r)\,\frac{\partial}{\sin\theta\partial\varphi} &
h_2^{\ell m}(t,r)\,\frac{\hat{X}_{\ell m}}{2\sin\theta}&
h_2^{\ell m}(t,r)\,\sin\theta\,\frac{\hat{W}_{\ell m}}{2}\\
h_0^{\ell m}(t,r)\,\frac{\sin\theta\partial}{\partial\theta} &
h_1^{\ell m}(t,r)\,\frac{\sin\theta\partial}{\partial\theta} &
h_2^{\ell m}(t,r)\,\sin\theta\,\frac{\hat{W}_{\ell m}}{2} &
-h_2^{\ell m}(t,r)\,\sin\theta\,\frac{\hat{X}_{\ell m}}{2}\\
\end{vmatrix}
Y_{\ell m}(\theta,\varphi)
\end{equation}
for the {\it odd} parity modes. Where $g_{\mu\nu}=g_{\mu\nu}^{\rm Schw}+h_{\mu\nu}$.

Above we used Zerilli's notation
\begin{equation}
\hat{X}_{\ell m}\dot=2\frac{\partial}{\partial\varphi}
\left(\frac{\partial}{\partial\theta}-\cot\theta\right),
\end{equation}
\begin{equation}
\hat{W}_{\ell m}\dot=\left(\frac{\partial^2}{\partial\theta^2}
-\cot\theta\frac{\partial}{\partial\theta}
-\frac{1}{\sin^2\theta}\frac{\partial^2}{\partial\varphi^2}\right),
\end{equation}
and
\begin{equation}
\hat{Z}_{\ell m}\dot=\left(
\frac{\partial^2}{\partial\varphi^2}
+\sin\theta\cos\theta\frac{\partial}{\partial\theta}
\right).
\end{equation}
We will also introduce Zerilli's $\lambda$
\begin{equation}
 \lambda=(\ell-1)(\ell+2)/2.
\end{equation}

\subsection{Even Parity}

The Zerilli's (C7a)-(C7g) equations with corrections are
\begin{eqnarray}
& &\left(1-\frac{2M}{r}\right)^2\frac{\partial^2 K^{\ell m}}{\partial r^2}
+\frac{1}{r}\left(1-\frac{2M}{r}\right)
\left(3-\frac{5M}{r}\right)\frac{\partial K^{\ell m}}{\partial r}
\nonumber \\
& &-\frac{1}{r}\left(1-\frac{2M}{r}\right)^2
\frac{\partial H_2^{\ell m}}{\partial r}
-\frac{1}{r^2}\left(1-\frac{2M}{r}\right)(H_2^{\ell m}-K^{\ell m})\nonumber\\
& &-\frac{(\lambda+1)}{r^2}\left(1-\frac{2M}{r}\right)(H_2^{\ell m}+K^{\ell m})
= -8\pi A_{\ell m}^{(0)}\,, \label{C7a}
\end{eqnarray}
\begin{eqnarray}
& &\frac{\partial}{\partial t}\left[
\frac{\partial K^{\ell m}}{\partial r}+\frac{1}{r}(K^{\ell m}-H_2^{\ell m})
-\frac{M}{r(r-2M)}K^{\ell m} \right]
-\frac{(\lambda+1)}{r^2}H_1^{\ell m}\nonumber \\
&&=-4\sqrt{2}\pi iA_{\ell m}^{(1)}\,, \label{C7b}
\end{eqnarray}
\begin{eqnarray}
&&\left(\frac{r}{r-2M}\right)^2\frac{\partial^2 K^{\ell m}}{\partial t^2}
-\frac{r-M}{r(r-2M)}\frac{\partial K^{\ell m}}{\partial r}
-\frac{2}{r-2M}\frac{\partial H_1^{\ell m}}{\partial t}
\nonumber \\
&&+\frac{1}{r}\frac{\partial H_0^{\ell m}}{\partial r}
+\frac{1}{r(r-2M)}(H_2^{\ell m}-K^{\ell m})
\nonumber \\
&&+\frac{(\lambda+1)}{r(r-2M)}(K^{\ell m}-H_0^{\ell m}) = -8\pi A_{\ell m}\,,\label{C7c}
\end{eqnarray}
\begin{eqnarray}
&&\frac{\partial}{\partial r}\left[
\left(1-\frac{2M}{r}\right)H_1^{\ell m}\right]
-\frac{\partial}{\partial t}(H_2^{\ell m}+K^{\ell m})
=\frac{8\pi ir}{\sqrt{\lambda+1}} B_{\ell m}^{(0)}\,, \label{C7d}
\end{eqnarray}
\begin{eqnarray}
&&-\frac{\partial H_1^{\ell m}}{\partial t}+\left(1-\frac{2M}{r}\right)
\frac{\partial}{\partial r}(H_0^{\ell m}-K^{\ell m})+\frac{2M}{r^2}H_0^{\ell m}
\nonumber \\
&&+\frac{1}{r}\left(1-\frac{M}{r}\right)(H_2^{\ell m}-H_0^{\ell m})
=\frac{8\pi(r-2M)}{\sqrt{\lambda+1}}B_{\ell m}
\,, \label{C7e}
\end{eqnarray}
\begin{eqnarray}
&&-\frac{r}{r-2M}\frac{\partial^2 K^{\ell m}}{\partial t^2}
+\left(1-\frac{2M}{r}\right)\frac{\partial^2 K^{\ell m}}{\partial r^2}
+\frac{2}{r}\left(1-\frac{M}{r}\right)
\frac{\partial K^{\ell m}}{\partial r}
\nonumber \\
&&-\frac{r}{r-2M}\frac{\partial^2 H_2^{\ell m}}{\partial t^2}
+2\frac{\partial^2 H_1^{\ell m}}{\partial t\partial r}
-\left(1-\frac{2M}{r}\right)\frac{\partial^2 H_0^{\ell m}}{\partial r^2}
\nonumber \\
&&+\frac{2(r-M)}{r(r-2M)}\frac{\partial H_1^{\ell m}}{\partial t}
-\frac{1}{r}\left(1-\frac{M}{r}\right)
\frac{\partial H_2^{\ell m}}{\partial r}
-\frac{r+M}{r^2}\frac{\partial H_0^{\ell m}}{\partial r}
\nonumber \\
&&+\frac{(\lambda+1)}{r^2}(H_0^{\ell m}-H_2^{\ell m}) =
8\sqrt{2}\pi G_{\ell m}^{(s)}\,, \label{C7f}
\end{eqnarray}
\begin{eqnarray}
&&\frac{H_0^{\ell m}-H_2^{\ell m}}{2} =
\frac{8\pi r^2F_{\ell m}}{\sqrt{2\lambda(\lambda+1)}}\,.\label{C7g}
\end{eqnarray}


\subsection{Odd Parity}
The Einstein equation in the Regge-Wheeler { gauge} for the odd
parity sector are [See Zerilli's\cite{Zerilli70a} equations (C6a)-(C6c)).
Note the corrections to the source terms.]
\begin{eqnarray}
&&\frac{\partial^2h_0^{\ell m}}{\partial r^2}-\frac{\partial^2h_1^{\ell m}}
{\partial r\partial t} -\frac2r\frac{\partial h_1^{\ell m}}{\partial t}+
\left[\frac{4M}{r^2}-\frac{2(\lambda+1)}{r}\right]
\frac{h_0^{\ell m}}{r-2M}\nonumber\\
&=&-\frac{8\pi\,rQ^{(0)}_{\ell m}}{(1-\frac{2M}{r})\sqrt{(\lambda+1)}},\label{tphiodd}
\end{eqnarray}
\begin{eqnarray}
&&\frac{\partial^2h_1^{\ell m}}{\partial t^2}-\frac{\partial^2h_0^{\ell m}}{\partial
r\partial t} +\frac2r\frac{\partial h_0^{\ell m}}{\partial t}+2\lambda
(r-2M)\frac{h_1^{\ell m}}{r^3}=\frac{8\pi\,i(r-2M)Q_{\ell m}}{\sqrt{(\lambda+1)}},\label{rphiodd}
\end{eqnarray}
\begin{eqnarray}
&&(1-\frac{2M}{r})\frac{\partial h_1^{\ell m}}{\partial r}-\frac{1}{(1-\frac{2M}{r})}
\frac{\partial h_0^{\ell m}}{\partial t}
+\frac{2M}{r^2}h_1^{\ell m}=-\frac{4\pi\,ir^2D_{\ell m}}{\sqrt{\lambda(\lambda+1)}},
\label{thetaphiodd}
\end{eqnarray}
where $Q^{(0)}_{\ell m}, Q_{\ell m}$ and $D_{\ell m}$ give the multipole
decomposition of the energy-momentum tensor (See Table~\ref{Tmunu}).


\subsection{Source terms}

Table \ref{Tmunu} gives the source terms produced by an orbiting
particle in the Schwarzschild background after decomposition of the
Stress-Energy Tensor into tensor harmonics.  There, $U^0(t)=dt/d\tau$,
is the zeroth component of the four-velocity of the particle and
$\Omega_p(t)$ is its angular location.

\bigskip
\tablecaption{Energy-momentum-Stress Tensor in terms of Tensor Harmonics}\label{Tmunu}
\renewcommand{\arraystretch}{1.5}
\begin{supertabular}{|c|}
\hline\hline
$A_{\ell m}(r,t)=m_0 U^0(t) \left({dR \over dt}\right)^2(r-2M)^{-2}
\delta[r-R(t)]\overline{Y}_{\ell m}(\Omega_p(t))$
\\ 
\hline
$\displaystyle
A_{\ell m}^{(0)}=m_0 U^0(t) \left(1-{2M \over r}\right)^2r^{-2}
\delta[r-R(t)]\overline{Y}_{\ell m}(\Omega_p(t))$
\\ 
\hline
$\displaystyle
A_{\ell m}^{(1)}=\sqrt{2}im_0 U^0(t) {dR \over dt}r^{-2}
\delta[r-R(t)]\overline{Y}_{\ell m}(\Omega_p(t))$
\\ 
\hline
 $\displaystyle
B_{\ell m}^{(0)}=[\lambda+1]^{-1/2}im_0 U^0(t)
\left(1-{2M \over r}\right)r^{-1}
\delta[r-R(t)]\frac{d\overline{Y}_{\ell m}}{dt}(\Omega_p(t))$
\\
\hline
 $\displaystyle
B_{\ell m}=[\lambda+1]^{-1/2}m_0 U^0(t)
(r-2M)^{-1}{dR \over dt}
\delta[r-R(t)]\frac{d\overline{Y}_{\ell m}}{dt}(\Omega_p(t))$
\\ 
\hline
 $\displaystyle
Q_{\ell m}^{(0)}=[\lambda+1]^{-1/2}m_0 U^0(t)
\left(1-{2M \over r}\right)r^{-1}
\delta[r-R(t)]
$
\\
$\displaystyle
\times \left[{1\over \sin \Theta}{\partial \overline{Y}_{\ell m} \over \partial \Phi}
{d \Theta \over dt}-\sin \Theta{\partial \overline{Y}_{\ell m} \over \partial \Theta}
{d\Phi \over dt}\right]$
$$ \\ \hline
$\displaystyle
Q_{\ell m}=[\lambda+1]^{-1/2}im_0 U^0(t) {dR \over dt}
(r-2M)^{-1}
\delta[r-R(t)]
$
\\
$\displaystyle
\times \left[{1\over \sin\Theta}{\partial \overline{Y}_{\ell m} \over \partial \Phi}
{d \Theta \over dt}-\sin \Theta{\partial \overline{Y}_{\ell m} \over \partial \Theta}
{d\Phi \over dt}\right]$
$$ \\ \hline
$\displaystyle
D_{\ell m}=-[2\lambda(\lambda+1)]^{-1/2}im_0 U^0(t)
\delta[r-R(t)]$
\\
$\displaystyle
\times \left({1 \over 2}\left[({d \Theta \over dt})^2
-\sin^2\Theta ({d\Phi \over dt})^2\right]
{1 \over \sin \Theta}\overline{X}_{\ell m}[\Omega(t)]
-\sin \Theta {d\Phi \over dt}{d \Theta \over dt}
  \overline{W}_{\ell m}[\Omega(t)]\right)$
$$ \\ \hline
$\displaystyle
F_{\ell m}=[2\lambda(\lambda+1)]^{-1/2}m_0 U^0(t)
\delta[r-R(t)]$
\\
$\displaystyle
\times \left({d\Phi \over dt}{d \Theta \over dt}\overline{X}_{\ell m}[\Omega(t)]
+{1 \over 2}\left[({d \Theta \over dt})^2
-\sin^2\Theta ({d\Phi \over dt})^2\right]\overline{W}_{\ell m}[\Omega(t)]\right)$
$$\\ \hline
 $\displaystyle
G_{\ell m}^{(s)}={m_0 U^0(t) \over \sqrt{2}}\delta[r-R(t)]
\left[({d \Theta \over dt})^2
+\sin^2\Theta ({d\Phi \over dt})^2\right]\overline{Y}_{\ell m}(\Omega_p(t))$
\\
\hline\hline
\end{supertabular}

\section{Reconstruction in terms of metric perturbations waveforms}\label{Recon}

Here we recall the metric reconstruction in the original Schwarzschild
perturbations approach based on waveforms for the even and odd parity
perturbations (Zerilli's and Regge-Wheeler respectively). We first introduce
the gauge invariant expressions for these waveforms. We then make use of
the general relativistic field equations, in the Regge-Wheeler gauge, to
solve for the metric perturbations, including nonvanishing matter terms.

\subsection{Even Parity}

We consider the following waveform\cite{Lousto97b} in terms of
generic metric perturbations in the Regge-Wheeler notation

\begin{eqnarray}\label{psieven}
\psi_{\rm even}^{\ell m}(r,t)&=&\frac{r}{(\lambda+1)}\left[K^{\ell m}+
\frac{r-2M}{\lambda r+3M}\left(H_2^{\ell m}-r\partial_rK^{\ell m}\right)\right]
\nonumber\\
&+&\frac{r-2M}{\lambda r+3M}\left[r^2\partial_rG^{\ell m}-2h_1^{\ell m}\right],
\end{eqnarray}

This is related to Zerilli's\cite{Zerilli70a} even parity waveforms
$\psi_{Zer}^{\ell m}$ by
\begin{equation}
\partial_t\psi_{\rm even}^{\ell m}=\psi_{Zer}^{\ell m}
-\,{\frac {4\pi i\,\sqrt {2}\,{r}^{2}\left (r-2\,M
\right ){ A^{(1)}_{\ell m}}}{\left (\lambda+1\right )
\left (\lambda\,r+3\,M\right )}},
\end{equation}
where for an orbiting particle\cite{Zerilli70a}
\begin{equation}
A^{(1)}_{\ell m}=im_0\sqrt{2}\left(\frac{U^0(t)}{r^2}\right)
\left(\frac{dR}{dt}\right)\overline{Y}_{\ell m}\delta[r-R(t)],
\end{equation}
and it relates to Moncrief's\cite{Moncrief74} waveform
$\psi_{Mon}^{\ell m}$ by
\begin{equation}
\psi_{\rm even}^{\ell m}=\frac{\psi_{Mon}^{\ell m}}{(\lambda+1)}.
\end{equation}

The $tt$ component of Hilbert-Einstein's equations gives us the
Hamiltonian constraint. In the Regge-Wheeler gauge $(h_1^{\ell m}=h_0^{\ell m}=G^{\ell m}=0)$ it
is given by 
Eq.\ (\ref{C7a}).  Only two metric
coefficients $(K^{\ell m}$ and $H_2^{\ell m})$ appear in this equation and
none of its time derivatives.
Considering the Regge-Wheeler gauge, the definition of $\psi_{\rm
even}^{\ell m} $ (see Eq.~(\ref{psieven}), and the Hamiltonian
constraint, Eq.~(\ref{C7a}), we can express these two metric
coefficients in terms of $\psi_{\rm even}^{\ell m}$ (and source
terms) only
\begin{eqnarray}\label{Krw}
K^{\ell m}&=&\frac{6M^2+3M\lambda r+\lambda (\lambda +1)r^2} {r^2(\lambda
r+3M)}\psi_{\rm even}^{\ell m} +\left( 1-\frac{2M}r\right) \,\partial_r
\psi_{\rm even}^{\ell m}\nonumber\\ 
&&-\frac{8\pi r^3 A^{(0)}_{\ell m}}{(\lambda +1)(\lambda r+3M)},
\end{eqnarray}
and
\begin{eqnarray} \label{H2}
&&H_2^{\ell m}=-\frac{9M^3+9\lambda M^2r+ 3\lambda ^2Mr^2+\lambda
^2(\lambda +1)r^3}{ r^2(\lambda r+3M)^2}\,\psi_{\rm even}^{\ell m} \nonumber\\
&&+\frac{3M^2-\lambda Mr+\lambda r^2}{r(\lambda r+3M)}\partial
_r\psi_{\rm even}^{\ell m} +(r-2M)\partial _r^2\psi_{\rm even}^{\ell m} \nonumber \\ &&-\frac{8\pi r^4}{(\lambda +1)(\lambda r+3M)}\partial_r A^{(0)}_{\ell m}\nonumber\\
&&+\frac{8\pi r^3({\lambda}^{2}{r}^{2}-2\,\lambda\,{r}^{2}+10\,\lambda\,rM-9\,rM+27\,{M}^{2})}{(\lambda +1)(r-2M)(\lambda r+3M)^2} A^{(0)}_{\ell m}.
\end{eqnarray}

From 
Eq.~(\ref{C7b}) and the expressions for
$\partial_t K^{\ell m}$ and $\partial_t{H}_2^{\ell m}$ in terms of
$\partial_t{\psi_{\rm even}^{\ell m} }$, we find the $H_1^{\ell m}$ metric
coefficient in the Regge-Wheeler gauge
\begin{eqnarray}\label{H1}
&&H_1^{\ell m} =r\partial_r(\partial_t{\psi_{\rm even}^{\ell m} })+
\frac{\lambda r^2-3M\lambda r-3M^2}{\left( r-2M\right)
(\lambda r+3M)}\partial_t\psi_{\rm even}^{\ell m} \nonumber\\
&&-\frac{8\pi r^5}{(\lambda +1)(r-2M)(\lambda r+3M)}\partial_t A^{(0)}_{\ell m}
+\frac{4\sqrt{2}i\pi r^2}{(\lambda +1)} A^{(1)}_{\ell m}.
\end{eqnarray}

These equations together with
\begin{eqnarray}\label{H0}
H_0^{\ell m}=H_2^{\ell m}+\frac{16\pi r^2\,F_{\ell m}}{\sqrt{2\lambda(\lambda+1)}},
\end{eqnarray}
give us all metric perturbations on the
chosen hypersurface in terms only of $\psi_{\rm even}^{\ell m} $ and
$\partial_t{\psi_{\rm even}^{\ell m} }$ (and the source).
(See also Ref.~\cite{Jhingan:2002kb} for general source expressions)

For the specific case of interest of a pointlike particle we have
\begin{eqnarray}\label{Kp}
K^{\ell m}&=&\frac{6M^2+3M\lambda r+\lambda (\lambda +1)r^2} {r^2(\lambda
r+3M)}\psi_{\rm even}^{\ell m} +\left( 1-\frac{2M}r\right) \,\partial_r
\psi_{\rm even}^{\ell m}\nonumber\\ &&-\frac{8\pi m_0 \overline{Y}_{\ell m}(t) \
U^0(t)(r-2M)^2}{(\lambda +1)(\lambda r+3M)r}\delta [r-R(t)]\ .
\end{eqnarray}

\begin{eqnarray} \label{H2p}
&&H_2^{\ell m}=-\frac{9M^3+9\lambda M^2r+ 3\lambda ^2Mr^2+\lambda
^2(\lambda +1)r^3}{ r^2(\lambda r+3M)^2}\,\psi_{\rm even}^{\ell m} \nonumber\\
&&+\frac{3M^2-\lambda Mr+\lambda r^2}{r(\lambda r+3M)}\partial
_r\psi_{\rm even}^{\ell m} +(r-2M)\partial _r^2\psi_{\rm even}^{\ell m} \nonumber \\ &&+\frac{8\pi
m_0 \overline{Y}_{\ell m}(t) U^0(t)(1-\frac{2M}{r})[\lambda ^2r^2+2\lambda
Mr-3Mr+3M^2]}{(\lambda +1)(\lambda r+3M)^2}\delta [r-R(t)]\nonumber\\
&&-\frac{8\pi m_0 \overline{Y}_{\ell m}(t) U^0(t)(r-2M)^2}{ (\lambda
+1)(\lambda r+3M)}\delta' [r-R(t)]\ .
\end{eqnarray}

\begin{eqnarray}\label{H1p}
&&H_1^{\ell m} =r\partial _r(\partial_t{\psi_{\rm even}^{\ell m} })+\frac{\lambda r^2-3M\lambda
r-3M^2}{%
\left( r-2M\right) (\lambda r+3M)}\partial_t{\psi_{\rm even}^{\ell m} }\nonumber\\
&&-\frac{8\pi m_0 \overline{Y}_{\ell m}(t) \ U^0(t)\stackrel{.}{r}_p(\lambda
r+M)} {(\lambda +1)(\lambda r+3M)}\delta [r-R(t)]\nonumber\\
&&+\frac{8\pi m_0
\overline{Y}_{\ell m}(t) \ U^0(t)\stackrel{.}{r}_pr(r-2M)}{(\lambda
+1)(\lambda r+3M)}\delta' [r-R(t)]. \nonumber\\
&&-\frac{8\pi
m_0(d\overline{Y}_{\ell m}/dt) (r-2M) r U^0(t)}{(\lambda +1)(\lambda r+3M)}
\delta[r-R(t)]\ .
\end{eqnarray}
and
\begin{eqnarray}\label{H0p}
H_0^{\ell m}=H_2^{\ell m}+16\pi r^2\,m_0\,U^0(t)\,{\rm ang1}(t)\,\delta[r-R(t)],
\end{eqnarray}
where
\begin{eqnarray}\label{ang1}
{\rm ang1}(t)&=&\frac12\left[\left(\frac{d\Theta}{dt}\right)^2
-\sin^2\Theta\left(\frac{d\Phi}{dt}\right)^2\right]\overline{W}^{\ell m}
+\frac{d\Phi}{dt}\frac{d\Theta}{dt}\overline{X}^{\ell m},
\end{eqnarray}
\begin{eqnarray}
\overline{X}^{\ell m}&=&2\partial_\varphi\Big(\partial_\theta-\cot\theta\Big)
\overline{Y}^{\ell m},\\
\nonumber\\
\overline{W}^{\ell m}&=&\left(\partial^2_\theta-\cot\theta\,\partial_\theta-
\frac{1}{\sin^2\theta}\partial^2_\varphi\right)\overline{Y}^{\ell m}.
\end{eqnarray}

\subsection{Odd Parity}

We consider the following waveform in terms of generic metric
perturbations in the Regge-Wheeler notation

\begin{eqnarray}\label{psiodd}
\psi^{\ell m}_{\rm odd}(r,t)&=&\frac{r}{\lambda}
\left[r^2\partial_r\left(\frac{h_0^{\ell m}(r,t)}{r^2}\right)-
\partial_t h_1^{\ell m}(r,t)\right]\nonumber\\
&=&\frac{2r}{\lambda}\sqrt{1-\frac{2M}{r}}K_{r\theta}.
\end{eqnarray}

This waveform is related to the Zerilli's\cite{Zerilli70a} and
Moncrief's\cite{Moncrief74} odd parity waveforms
$\psi_{Zer}^{odd}=\psi_{Mon}^{odd}$
\begin{equation}
\psi_{Zer}^{odd}=\frac{(1-\frac{2M}{r})}{r}\left[h_1^{\ell m}
+\frac{r^2}{2}\partial_r\left(\frac{h_2^{\ell m}}{r^2}\right)\right],
\end{equation}
by (See Eq.~(\ref{rphiodd}))
\begin{equation}
\partial_t\psi_{\rm odd}^{\ell m}=2\psi_{Zer}^{odd}
-\frac{8\pi\,i\,r(r-2M)Q^{\ell m}}{\lambda\sqrt{\lambda+1}},
\end{equation}
to the Cunningham et al \cite{Cunningham78} waveform $\psi_G^{\ell m}$ by 
\begin{equation}
\psi_{\rm odd}^{\ell m}=-2\frac{(\ell-2)!}{(\ell+2)!}\psi_G^{\ell m}
=-\frac{1}{2}\frac{\psi_G^{\ell m}}{\lambda(\lambda+1)},
\end{equation}
and  to the Weyl scalar $\Psi_2$
\begin{equation}
\Psi_2^-=\frac{(\ell+2)!}{8(\ell-2)!}\frac{\psi_{\rm odd}^{\ell m}}{r^3}.
\end{equation}
[Here we used the Kinnersley tetrad, in the Schwarzschild background,
and decomposed $\Psi_2$ into spherical harmonics].

One can use the field equations to write the metric perturbation in the
Regge--Wheeler { gauge}
\begin{eqnarray}
h_0^{\ell m}(r,t)&=&\frac12(1-\frac{2M}{r})\partial_r\left(r\psi^{\ell
m}_{\rm odd}\right)\label{h0odd}\label{h0Q}+\frac{4\pi r^3Q^{(0)}_{\ell
m}}{\lambda\sqrt{(\lambda+1)}}\\
\nonumber\\
h_1^{\ell m}(r,t)&=&\frac12\frac{r}{(1-\frac{2M}{r})}\partial_t\psi^{\ell
m}_{\rm odd}\label{h1odd}\label{h1Q}+\frac{4\pi i r^3 Q_{\ell
m}}{\lambda\sqrt{(\lambda+1)}}\nonumber
\end{eqnarray}

For a source term represented by a particle
the corresponding metric perturbations in the
Regge-Wheeler { gauge} are
\begin{eqnarray}
h_0^{\ell m}(r,t)&=&\frac12(1-\frac{2M}{r})\partial_r\left(r\psi^{\ell
m}_{\rm odd}\right)\\ &+&\frac{4\pi m_0 r(r-2M)U^0(t){\rm
ang}(t)\delta[r-R(t)]}{\lambda(\lambda+1)}\nonumber\\
\nonumber\\
h_1^{\ell m}(r,t)&=&\frac12\frac{r}{(1-\frac{2M}{r})}\partial_t\psi^{\ell m}_{\rm odd}\\
&-&\frac{4\pi m_0 r^3 U^0(t)(\frac{d}{dt}R){\rm
ang}(t)\delta[r-R(t)]}{(r-2M)\lambda(\lambda+1)},\nonumber
\end{eqnarray}
where
\begin{eqnarray}
{\rm ang}(t)&=&\frac{1}{\sin\Theta}\left(\frac{d\Theta}{dt}\right)
\partial_\varphi\overline{Y}^{\ell m}(\Theta,\Phi)
-\sin\Theta\left(\frac{d\Phi}{dt}\right)
\partial_\theta\overline{Y}^{\ell m}(\Theta,\Phi),
\end{eqnarray}
and $R(t),\Theta,\Phi$ define the trajectory of the orbiting particle 
in spherical coordinates.


\section{(Inverse) Chandrasekhar transformations in the time domain}\label{Chandra}

Chandrasekhar transformations deal with the expressions that relate the
waveforms in the metric perturbation picture (Regge-Wheeler and
Zerilli's) and the Weyl scalars that describe the curvature
perturbations (Teukolsky's). Chandrasekhar \cite{Chandrasekhar83}
found the transformations in the frequency domain. Those expressions
can be generalized to the time domain to describe local
transformations, and take into account matter terms such as a
particle. Below we give the explicit expressions for the
transformations (we drop the $(\ell m)$ superscript in the waveforms
for the sake of notational simplicity.)

\subsection{From Waveforms to Weyl scalars}


To obtain the Weyl scalars from the waveforms (\ref{psieven}) and
(\ref{psiodd}) for even and odd parity respectively we simply
substitute into Eqs.~(\ref{RWSpsi4+}), (\ref{RWSpsi0+}), (\ref{RWpsi4}), and
(\ref{RWpsi0}) the expressions (\ref{Krw})--(\ref{H0}), and
(\ref{h0Q})--(\ref{h1Q})
for the metric coefficients in the Regge-Wheeler gauge. The result is
\begin{eqnarray}
\psi_4^+&=&\frac{1}{16r}\sqrt{\frac{(\ell-2)!}{(\ell+2)!}}
\left\{2\psi_{,r*r*}^{\rm even}-2\psi_{,tr*}^{\rm even}+W^+
(\psi_{,r*}^{\rm even}-\psi_{,t}^{\rm even})-V^+\psi^{\rm even}\right.\nonumber\\&&
\left.+\frac{16\pi r^3}{(\lambda r+3M)(\lambda+1)}\left(\partial_t A^{(0)}_{\ell m}
-\partial_{r*} A^{(0)}_{\ell m}\right)
-{\frac {8\,i \left( r-2\,M \right) \sqrt {2}\pi \,A^{(1)}_{\ell m} \left( r,t
 \right) }{\lambda+1}}\right.\nonumber\\&&
\left.
+16\,{\frac {\pi \,r \left( {\lambda}^{2}{r}^{2}
-2\,\lambda\,{r}^{2}+10\,\lambda\,rM-9\,rM+27\,{M}^{2} \right)\,A^{(0)}_{\ell m} \left( r,t \right) }{ \left( \lambda+1 \right)  \left( \lambda\,r+3
\,M \right) ^{2}}}\right.\nonumber\\&&
\left.
-8\,{\frac {F_{\ell m} \left( r,t \right) \sqrt {2}\pi \,
 \left( r-2\,M \right) }{\sqrt {\lambda\, \left( \lambda+1 \right) }}}
\right\},
\end{eqnarray}
\begin{eqnarray}
\psi_4^-&=&\frac{-i}{16r}\sqrt{\frac{(\ell-2)!}{(\ell+2)!}}
\left\{2\psi_{,r*r*}^{\rm odd}-2\psi_{,tr*}^{\rm odd}+W^-
(\psi_{,r*}^{\rm odd}-\psi_{,t}^{\rm odd})-V^-\psi^{\rm odd}\right.\nonumber\\&&
\left.-\frac{16\pi r^2}{\lambda(\lambda+1)}\left(\partial_t Q^{(0)}_{\ell m}
-\partial_{r*} Q^{(0)}_{\ell m}\right)
+\frac{16i\pi r(r-2M)}{\lambda(\lambda+1)}\left(\partial_t Q_{\ell m}
-\partial_{r*} Q_{\ell m}\right)\right.\nonumber\\&&
\left.-{\frac {48\,i\pi \, \left( r-2\,M \right) ^{2}Q_{\ell m} }{
\lambda\,\sqrt {\lambda+1}r}}+{\frac {16\pi \, \left( 3\,r-8\,M
 \right) Q^{(0)}_{\ell m}  }{\lambda\,\sqrt {\lambda+1}}}-S^-
\right\},
\end{eqnarray}
\begin{eqnarray}
\psi_0^+&=&\frac{1}{4f^2r}\sqrt{\frac{(\ell-2)!}{(\ell+2)!}}
\left\{2\psi_{,r*r*}^{\rm even}+2\psi_{,tr*}^{\rm even}+W^+
(\psi_{,r*}^{\rm even}+\psi_{,t}^{\rm even})-V^+\psi^{\rm even}\right.\nonumber\\&&
\left.-\frac{16\pi r^3}{(\lambda r+3M)(\lambda+1)}\left(\partial_t A^{(0)}_{\ell m}
+\partial_{r*} A^{(0)}_{\ell m}\right)
+{\frac {8\,i \left( r-2\,M \right) \sqrt {2}\pi \,A^{(1)}_{\ell m} \left( r,t
 \right) }{\lambda+1}}\right.\nonumber\\&&
\left.
+16\,{\frac {\pi \,r \left( {\lambda}^{2}{r}^{2}
-2\,\lambda\,{r}^{2}+10\,\lambda\,rM-9\,rM+27\,{M}^{2} \right)\,A^{(0)}_{\ell m} \left( r,t \right) }{ \left( \lambda+1 \right)  \left( \lambda\,r+3
\,M \right) ^{2}}}\right.\nonumber\\&&
\left.
-8\,{\frac {F_{\ell m} \left( r,t \right) \sqrt {2}\pi \,
 \left( r-2\,M \right) }{\sqrt {\lambda\, \left( \lambda+1 \right) }}}
\right\},
\end{eqnarray}
\begin{eqnarray}
\psi_0^-&=&\frac{i}{4f^2r}\sqrt{\frac{(\ell-2)!}{(\ell+2)!}}
\left\{2\psi_{,r*r*}^{\rm odd}+2\psi_{,tr*}^{\rm odd}+W^-
(\psi_{,r*}^{\rm odd}+\psi_{,t}^{\rm odd})-V^-\psi^{\rm odd}\right.\nonumber\\&&
\left.\frac{16\pi r^2}{\lambda(\lambda+1)}\left(\partial_t Q^{(0)}_{\ell m}
+\partial_{r*} Q^{(0)}_{\ell m}\right)
+\frac{16i\pi r(r-2M)}{\lambda(\lambda+1)}\left(\partial_t Q_{\ell m}
+\partial_{r*} Q_{\ell m}\right)\right.\nonumber\\&&
\left.+{\frac {48\,i\pi \, \left( r-2\,M \right) ^{2}Q_{\ell m} }{
\lambda\,\sqrt {\lambda+1}r}}+{\frac {16\pi \, \left( 3\,r-8\,M
 \right) Q^{(0)}_{\ell m}  }{\lambda\,\sqrt {\lambda+1}}}-S^-
\right\},
\end{eqnarray}
where we introduced the Chandrasekhar notation for
\begin{eqnarray}
V^+&=&2\left(1-\frac{2M}{r}\right)
\frac{\left[\lambda^2(\lambda+1)r^3+3\lambda^2Mr^2+9\lambda M^2r+9M^3\right]}
{r^3(\lambda r+3M)^2 },\\
V^-&=&2\left(1-\frac{2M}{r}\right)\left(\frac{\lambda+1}{r^2}-3\frac{M}{r^3}\right),\\
W^+&=&2\frac{(\lambda r^2-3\lambda M r-3M^2)}{r^2(\lambda r+3M)},\\
W^-&=&2\frac{(r-3M)}{r^2}
\end{eqnarray}
and
\begin{eqnarray}
S^-&=& -{\frac {8\pi \,\left (r-2\,M\right
)}{\lambda\sqrt {\left (\lambda+1\right )}}} \left [{\frac {
\partial }{\partial r}}\left(r{Q^{(0)}}(r,t)\right)-ir{\frac {\partial }{
\partial t}}Q(r,t)
\right],
\end{eqnarray}
is the source term for the Regge-Wheeler wave equation.

These relations are local, and in the time domain. Compare to the
expressions in the frequency domain of Ref.~\cite{Chandrasekhar83},
Eqs. (345) and (353) in Chapter 4.

\subsection{From Weyl scalars to Waveforms}

To obtain the inverse Chandrasekhar relations we make use of
Eqs.~(\ref{H1psi})--(\ref{Kbis}), and (\ref{h0})--(\ref{h1}) for the
metric coefficients entering in the definitions of the even and odd
parity waveforms, Eqs.~(\ref{psieven}) and (\ref{psiodd}), respectively
\begin{eqnarray}
&&\psi^{\rm even}=
{\frac {{r}^{2} \left( r-2\,M \right) {\frac {\partial ^{2}}{\partial 
{r}^{2}}}{H_0} \left( r,t \right) }{\lambda\, \left( \lambda+1
 \right) }}-{\frac {{r}^{3}{\frac {\partial ^{2}}{\partial r\partial t
}}{H_1} \left( r,t \right) }{\lambda\, \left( \lambda+1 \right) }}\nonumber\\
&&+{\frac {r \left( rM\lambda-3\,{M}^{2}+{r}^{2}\lambda+6\,rM \right) {
\frac {\partial }{\partial r}}{H_0} \left( r,t \right) }{ \left( 
\lambda+1 \right) \lambda\, \left( \lambda\,r+3\,M \right) }}\nonumber\\
&&+{\frac {
 \left( 2\,{r}^{2}\lambda-5\,rM\lambda-21\,{M}^{2}+9\,rM \right) {r}^{
2}{\frac {\partial }{\partial t}}{H_1} \left( r,t \right) }{
 \left( \lambda+1 \right)  \left( \lambda\,r+3\,M \right)  \left( -r+2
\,M \right) \lambda}}\nonumber\\
&&-{\frac { \left( {M}^{2}\lambda\,r+2\,{r}^{2
}M{\lambda}^{2}-12\,{r}^{2}M-{r}^{2}M\lambda-2\,{r}^{3}\lambda+42\,r{M
}^{2}-{r}^{3}{\lambda}^{2}-63\,{M}^{3} \right) {H_0} \left( r,t
 \right) }{2 \left( \lambda+1 \right)  \left( \lambda\,r+3\,M \right) 
 \left( -r+2\,M \right) \lambda}}\nonumber\\
&&-4\,{\frac {{r}^{2}\pi \,\sqrt {2\,
\lambda+2} \left( -5\,{r}^{2}\lambda-12\,rM+9\,rM\lambda+21\,{M}^{2}
 \right) \sqrt {2}B \left( r,t \right) }{ \left( \lambda+1 \right) ^{2
}\lambda\, \left( \lambda\,r+3\,M \right) }}\nonumber\\
&&+{\frac {{r}^{4}{A^{(0)}}
 \left( r,t \right) }{\lambda\, \left( \lambda+1 \right)  \left( -r+2
\,M \right) }}\nonumber\\
&&+4\,{\frac {\sqrt {2}\sqrt {\lambda\, \left( \lambda+1
 \right) }{r}^{2}\pi \, \left( 2\,rM-11\,{M}^{2}-{r}^{2}\lambda+2\,rM
\lambda \right) F \left( r,t \right) }{ \left( -r+2\,M \right) 
 \left( \lambda+1 \right) ^{2}{\lambda}^{2}}}\nonumber\\
&&-8\,{\frac {{r}^{3}
 \left( -r+2\,M \right) \pi \,{\frac {\partial }{\partial r}}B \left( 
r,t \right) }{ \left( \lambda+1 \right) ^{3/2}\lambda}}-8\,{\frac {{r}
^{3}M\pi \,\sqrt {2}{\frac {\partial }{\partial r}}F \left( r,t
 \right) }{ \left( \lambda+1 \right) \sqrt {\lambda\, \left( \lambda+1
 \right) }\lambda}},
\end{eqnarray}
and 
\begin{eqnarray}
\psi^{\rm odd}&=&\frac{r}{\lambda}\left\{
-\frac{2}{r}\frac{(1-\frac{M}{r})}{(1-\frac{2M}{r}}h_0+S_0-\partial_th_1
\right\}.
\end{eqnarray}
So, finally
\begin{eqnarray}
\psi^{\rm odd}&=&\frac{r}{\lambda}\left\{
-\frac{2}{r}(1-\frac{M}{r})\int_{2M}^r\,\frac{S_0(r',t)}{1-\frac{2M}{r'}}\,dr'+
\frac{4ir^2}{f}\sqrt{\frac{(\ell+2)!}{(\ell-2)!}}
\left(\psi_4^--\frac{f^2}{4}\psi_0^-\right)\right.\nonumber\\
&&\left.-\frac{2}{\sqrt{1-\frac{2M}{r}}}\left[
\int_{2M}^r\,\partial_t S_1(r',t)\,\sqrt{1-\frac{2M}{r'}}\,dr'
\right]\right\}.
\end{eqnarray}


\section{Symmetric tetrad}\label{symtetrad}

A further algebraic simplification of the expressions can be achieved
by choosing the background tetrad such that we treat $\psi_4$ and
$\psi_0$ on the same footing, thus allowing simple linear combinations
of the sort $\psi^S_4\pm \psi^S_0$ in the expression in
Section~\ref{RWG}.  The components of the {\it symmetric} null tetrad
for the Kerr background are given by
\begin{subequations}\label{tetradS}
\begin{eqnarray}
   (\mathrm{l_S}^{\alpha}) &=& \left( \frac{r^2+a^2}{\sqrt{2\Delta\Sigma}},
	\sqrt{\frac{\Delta}{2\Sigma}},0,\frac{a}{\sqrt{2\Delta\Sigma}}\right), \\
   (\mathrm{n_S}^{\alpha}) &=& \left( \frac{r^2+a^2}{\sqrt{2\Delta\Sigma}},
	-\sqrt{\frac{\Delta}{2\Sigma}},0,\frac{a}{\sqrt{2\Delta\Sigma}}\right), \\
   (\mathrm{m_S}^{\alpha}) &=& \frac{1}{\sqrt{2}(r+ia\cos\theta)}\,
                  \left( ia\sin\theta,0,1,\frac{i}{\sin\theta} \right).\quad\quad
\end{eqnarray}
\end{subequations}

With the above choice of the tetrad the spin
coefficients are 
\begin{eqnarray}\label{spincoeffS}
\nu_S&=&,\quad\sigma_S=,\quad\lambda_S=,\quad\kappa_S=0,\nonumber\\
\pi_S&=&\pi_K=ia\rho_K^2\frac{\sin\theta}{\sqrt{2}},\nonumber\\
\tau_S&=&\tau_K=-ia\rho_K\overline{\rho}_K\frac{\sin\theta}{\sqrt{2}},\nonumber\\
\rho_S&=&\mu_S=\sqrt{\frac{\Delta}{2\Sigma}}\rho_K\nonumber\\
\epsilon_S&=&\frac{\left[M(r^2-a^2\cos^2\theta)-a^2r\sin^2\theta\right]}
{2\sqrt{2\Delta\Sigma^3}}\nonumber\\
\gamma_S&=&\epsilon_S-ia\cos\theta\sqrt{\frac{\Delta}{2\Sigma^3}}\nonumber\\
\alpha_S&=&\frac{\left[(r^2+a^2)\cos\theta-2iar\sin^2\theta\right]}
{2\sqrt{2}\sin\theta}\left(\rho_K^2\overline\rho_K\right)\nonumber\\
\beta_S&=&-\frac{\left[(r^2+a^2)\cot\theta\right]}
{2\sqrt{2}}\left(\overline\rho_K^2\rho_K\right),\quad
\alpha_S-\overline{\beta}_S=\alpha_K-\overline{\beta}_K,
\end{eqnarray}
where an overbar stands for complex conjugation.

The Weyl scalars computed with the Symmetric tetrad relate to those
computed with the Kinnersley tetrad as follows
\begin{eqnarray}\label{WeylS}
\psi_4^S&=&\left({\frac{2\Sigma}{\Delta}}\right)\psi_4^K,\quad
\psi_3^S=\sqrt{\frac{2\Sigma}{\Delta}}\psi_3^K,\quad
\psi_2^S=\psi_2^K,\nonumber\\
\psi_1^S&=&\sqrt{\frac{\Delta}{2\Sigma}}\psi_1^K,\quad
\psi_0^S=\left({\frac{\Delta}{2\Sigma}}\right)\psi_0^K.
\end{eqnarray}

\bigskip\bigskip
\noindent
{\bf References}
\bigskip

\providecommand{\bysame}{\leavevmode\hbox to3em{\hrulefill}\thinspace}
\providecommand{\MR}{\relax\ifhmode\unskip\space\fi MR }
\providecommand{\MRhref}[2]{%
  \href{http://www.ams.org/mathscinet-getitem?mr=#1}{#2}
}
\providecommand{\href}[2]{#2}



\end{document}